# Microscopic origin of electric-field-induced modulation of Curie temperature in cobalt

Fuyuki Ando[1], Kihiro T. Yamada[1], Tomohiro Koyama[2], Mio Ishibashi[1],

Yoichi Shiota[1], Takahiro Moriyama[1], Daichi Chiba[2], and Teruo Ono[1, 3*]

[1] *Institute for Chemical Research, Kyoto University, Gokasho, Uji, Kyoto, 611-0011, Japan.*

[2] *Department of Applied Physics, Faculty of Engineering, The University of Tokyo, Hongo 7-3-1, Bunkyo, Tokyo 113-8656, Japan.*

[3] *Center for Spintronics Research Network, Graduate School of Engineering Science, Osaka University, Machikaneyama 1-3, Toyonaka, Osaka 560-8531, Japan*

[*]E-mail: ono@scl.kyoto-u.ac.jp

**Abstract**

The Curie temperature $T_C$ is one of the most fundamental physical properties of ferromagnetic materials and can be described by Weiss molecular field theory with the exchange interaction of neighboring atoms. Recently, the electric-field-induced modulation of $T_C$ has been demonstrated in transition metals. This can be interpreted as indirect evidence for the electrical modulation of exchange coupling. However, the scenario has not yet been experimentally verified. Here, we demonstrate the electrical control of exchange coupling in cobalt film from direct magnetization measurements. We find that the reduction in magnetization with temperature, which is caused by thermal spin wave excitation and scales with Bloch's law, clearly depends on the applied electric field. Furthermore, we confirm that the correlation between the electric-field-induced modulation of $T_C$ and that of exchange coupling follows Weiss molecular field theory.

**PACS numbers**: 71.70.Gm, 75.70.Ak, 71.20.Be, 83.60.Np

The electrical control of magnetism, such as magnetic anisotropy or the Curie temperature $T_C$, in ferromagnetic metal/insulator structures provides power-efficient operation in spintronic devices [1–14]. The electric field effect in 3$d$-transition ferromagnetic metals is considered to be due to the change in electron occupancy in the $d$-orbital state. However, the modulation of $T_C$ by an electric field experimentally observed in a Pt/Co film at room temperature [7], where the increase of electron number tends to increase $T_C$, exhibited an opposite trend to the theoretical expectation of the Curie temperature Slater–Pauling curve [15]. Recently, *ab initio* calculations suggested that the change in $T_C$ in a Pt/Co system can be attributed to the modulation of the Heisenberg exchange coupling [16,17]. Although the modulation of the exchange stiffness at room temperature in an ultrathin Co [18] and CoFeB [19,20] has recently been reported, thermal effects such as thermal spin wave excitation inhibited the determination of the exchange coupling $J_{ex}$ at low temperature. Therefore, the mechanism of the electric field control of $T_C$ has not yet been experimentally clarified. Here, we report the $J_{ex}$ modulation in an ultrathin Co layer by the application of a gate electric field from direct magnetization measurements. Furthermore, we show that the correlation between the electric-field-induced modulation of $T_C$ and that of $J_{ex}$ follows Weiss molecular field theory of ferromagnetism.

A perpendicularly magnetized Pt/Co system [21] was used to observe the modulation of its ferromagnetism by an electric field [6]. Multilayers of Ta (3.3 nm)/Pt (1.8 nm)/Co (0.23 nm)/MgO (2.0 nm)/HfO$_2$ (40 nm)/Cr (3.0 nm)/Au (10 nm) were prepared on an intrinsic Si(001) substrate with a thermally oxidized Si layer (see Supplementary Information). A schematic of the measurement configuration and a microscope image of the device are shown in Figs. 1(a) and 1(b), respectively. An external magnetic field was applied in the direction normal to the film surface by a superconductive magnet, and the perpendicular component of the magnetization was measured by a superconducting quantum interference device. Here, the application of the positive gate voltage $V_G$ corresponds to the increase of the electron density at the top surface of the Co layer. In this capacitor structure, a $V_G$ of +10 V corresponds to an electric field ($E$) of +2.38 MVcm$^{-1}$, which induces an accumulation of 0.017 electrons per Co atom at the top surface of the Co layer.

Magnetization measurements were performed in the absence of an external magnetic field under the application of $V_G$ = 0, ±10, ±14, and 0 V. The perpendicular remanence magnetization $m_\perp$ was almost equal to the saturation magnetization because of the squareness of the $m$–$H$ loops in Pt/Co (see Supplementary Information). Figure 2(a) shows the temperature dependence of $m_\perp$ normalized by $m_\perp$ at 0 K (= $m_S$). The results in the range of 10–140 K scale well with Bloch's law [22,23], which describes that the reduction in $m_\perp$ with temperature can be attributed to the spin wave excitation:

$$\frac{m_\perp(T)}{m_S} = 1 - \frac{0.0587}{SQ} \cdot \left(\frac{k_B T}{2J_{ex}S}\right)^{\frac{3}{2}} - \Theta\left(T^{\frac{5}{2}}, T^{\frac{7}{2}}\right) \tag{1}$$

where $S$ is the spin angular momentum ($S = 0.899$ in Pt/Co system [24]), $Q$ is the number of atoms per unit cell ($Q = 4$ for a fcc lattice), and $k_B$ is the Boltzmann constant. Bloch's law generally assumes a three-dimensional (3D) model, whereas the spontaneous magnetization linearly decreases with temperature in a two-dimensional (2D) system [25]. For this device, the reduction in perpendicular remanence magnetization agrees with Bloch's law (3D) rather than the linear line (2D). This result suggests that the effective thickness of the ferromagnetic layer is thicker than the nominal Co thickness (0.23 nm) because the attached Pt layer has a proximity-induced moment. The additional term $\Theta = \left(\frac{0.0493}{SQ}\right)^{\frac{5}{3}} \cdot \left(\frac{k_B T}{2J_{ex}S}\right)^{\frac{5}{2}} + \left(\frac{0.0701}{SQ}\right)^{\frac{7}{3}} \cdot \left(\frac{k_B T}{2J_{ex}S}\right)^{\frac{7}{2}}$ corrects the effective mass of the magnon with large velocity, allowing us to extend the applicable fitting range to higher temperature. As shown in Fig. 2(b), the temperature dependence of $m_\perp/m_S$ clearly depends on $V_G$, which suggests that $J_{ex}$ is modified by the application of $V_G$. The electric field dependence of $S$ is negligible because the predicted modulation of $S$ at 0 K [15,16] is one-order smaller than the observed modulation. Thus, $J_{ex}$ can be determined by fitting the results with Bloch's law. The variation in $J_{ex}$ under the application of $V_G$ is shown in Fig. 2(c). Note that $J_{ex}$ is tuned reversibly and the final $J_{ex}$ ($V_G = 0$ V) after the application of several $V_G$ values becomes equal to the initial $J_{ex}$ ($V_G = 0$ V) without hysteresis. This implies that the modulation of $J_{ex}$ observed here is an intrinsic electric field effect, unlike the electrochemical reaction [26,27] that shows hysteretic behavior because of the thermal activation process. The obtained absolute value of $J_{ex} = (2.465 \pm 0.002) \times 10^{-22}$ J is one-order smaller than that of bulk Co ($J_{ex} = 4.5 \times 10^{-21}$ J) [28], which is consistent with the lower Curie temperature $T_C$ of $231 \pm 1$ K.

To accurately determine $T_C$ by preventing the formation of multi-domain states, the temperature dependence of $m_\perp$ per unit area $A$ was also measured under a perpendicular magnetic field of 200 Oe. As shown in Fig. 3(a), $m_\perp/A$ rapidly decreases above 200 K. There is no difference between the cooling and heating processes, which implies that the device does not form multi-domain structure. The reduction in $m_\perp/A$ around $T_C$ can be quantified by the critical exponent as $\beta$ follows:

$$m_\perp \propto \left(1 - T/T_C\right)^\beta \qquad (2)$$

where the value of $\beta$ depends on the dimensionality and represents the number of degrees of freedom for spin orientation. The values predicted for the 2D Ising [29] and 2D XY [30] models are $\beta = 0.125$ and 0.23, respectively, whereas those for the 3D Ising, 3D

XY and 3D Heisenberg model [31] are $\beta$ = 0.326, 0.345 and 0.365, respectively. The measurement results under a magnetic field of 200 Oe are fitted with Eq. (2) in the range of $0.4 < (1 - T/T_C)^\beta < 0.6$. From the fit to the data, $\beta$ = 0.317 ± 0.008 is obtained for all values of $V_G$, which also indicates that the Pt/Co is near the 3D system. The spontaneous magnetizations independently determined by the Arrott plot almost match the fitting curve, supporting the validity of the determination of $T_C$ from the temperature dependence of $m_\perp/A$. The measurement was performed under the application of $V_G$ = 0, ±5, ±10, and 0 V, and a clear difference can be seen as shown in Fig. 3(b). The variation in $T_C$ under the $V_G$ shows the reversible change as well as that of $J_{ex}$ as shown in Fig. 3(c). $J_{ex}$ normalized by $J_{ex}$ ($V_G$ = 0 V) and $T_C$ normalized by $T_C$ ($V_G$ = 0 V) are plotted as a function of $V_G$ in Fig. 4(a). The linear relationship between $J_{ex}(V)$ and $T_C(V)$ reveals that the modulation of $J_{ex}$ is the microscopic origin of the change in $T_C$. Figure 4(b) shows the correlation between $J_{ex}(V)$ and $T_C(V)$ investigated for three devices A, B, and C. The difference of $T_C$ among the devices can be attributed to the distribution of the Co layer thickness in the same wafer [32]. All the plots modulated by $V_G$ are on a linear fitting line, which follows Weiss molecular field theory $T_C(V) \propto J_{ex}(V)$.

Finally, we discuss the mechanism of the observed electrical modulation of $J_{ex}$. Assuming we simply consider the change in the electron number at the top surface of the Co, the decrease (increase) in $J_{ex}$ as well as $T_C$ with increasing (decreasing) electron number is expected from the theoretical prediction of the Curie temperature Slater–Pauling curve [15]. This trend is inconsistent with our experimental results. Recently, *ab initio* calculations [16] suggested that the application of an electric field modifies not only the electron density at the surface of the Co layer on Pt(111) but also the *p*–*d* hybridization of Co. Even though the whole electron number in the Co atom increases under the application of a positive electric field, the *d* electron number decreases because of the change in *p*–*d* hybridization, which dominantly contributes to the modulation of $J_{ex}$. According to this consideration, the observed sign of the $J_{ex}$ modulation by an electric field is consistent with the Curie temperature Slater–Pauling curve. From the linear fit in Fig. 4(b), $D_0(V)/T_C(V) = 0.151 \pm 0.001$ meVÅ$^2$K$^{-1}$ is obtained, where $D_0(V) = 2J_{ex}(V)Sa^2$ is the spin wave stiffness at 0 K. This value is comparable to that of Co-alloys ($D_0(V)/T_C(V) \approx 0.22$ meVÅ$^2$K$^{-1}$ in Co-metalloid alloys and $\approx 0.3$ meVÅ$^2$K$^{-1}$ in Co-transition metal alloys) [33], suggesting that the electric field modulation of $J_{ex}$ and $T_C$ can be regarded as the perturbation by doping electrons to a rigid Co-3*d* band [15].

In summary, we experimentally demonstrated the electrical modulation of the exchange coupling in a Pt/Co structure from direct magnetization measurement. The reduction in perpendicular remanence magnetization with temperature scales well with Bloch's law for thermal spin wave excitation. We further provide a clear correlation between the electric-field-induced change in the Curie temperature and that in the exchange coupling, which follows Weiss molecular field theory. These results provide a deeper understanding of the microscopic origin of the electric field effect in ferromagnetic metals.

This work was partly supported by JSPS KAKENHI Grant Numbers 15H05702, 26870300, 25220604, 16H05977, and 26103002; the Collaborative Research Program of the Institute for Chemical Research, Kyoto University; the Cooperative Research Project Program of the Research Institute of Electrical Communication, Tohoku University; and the R&D project for ICT Key Technology of MEXT from the Japan Society for the Promotion of Science (JSPS).

**References**

[1]  H. Ohno, D. Chiba, and F. Matsukura, Nature **408**, 944 (2000).

[2]  H. Boukari, P. Kossacki, M. Bertolini, D. Ferrand, J. Cibert, S. Tatarenko, A. Wasiela, J. A. Gaj, and T. Dietl, Phys. Rev. Lett. **88**, 207204 (2002).

[3]  M. Weisheit, S. Fahler, A. Marty, Y. Souche, C. Poinsignon, and D. Givord, Science **315**, 349 (2007).

[4]  D. Chiba, M. Sawicki, Y. Nishitani, Y. Nakatani, F. Matsukura, and H. Ohno, Nature **455**, 515 (2008).

[5]  T. Maruyama, Y. Shiota, T. Nozaki, K. Ohta, N. Toda, M. Mizuguchi, A. A. Tulapurkar, T. Shinjo, M. Shiraishi, S. Mizukami, Y. Ando, and Y. Suzuki, Nat. Nanotechnol. **4**, 158 (2009).

[6]  M. Sawicki, D. Chiba, A. Korbecka, Y. Nishitani, J. a. Majewski, F. Matsukura, T. Dietl, and H. Ohno, Nat. Phys. **6**, 22 (2010).

[7]  D. Chiba, S. Fukami, K. Shimamura, N. Ishiwata, K. Kobayashi, and T. Ono, Nat. Mater. **10**, 853 (2011).

[8]  K. Shimamura, D. Chiba, S. Ono, S. Fukami, N. Ishiwata, M. Kawaguchi, K. Kobayashi, and T. Ono, Appl. Phys. Lett. **100**, 122402 (2012).

[9]  Y. Shiota, T. Nozaki, F. Bonell, S. Murakami, T. Shinjo, and Y. Suzuki, Nat. Mater. **11**, 39 (2012).

[10] W.-G. Wang, M. Li, S. Hageman, and C. L. Chien, Nat. Mater. **11**, 64 (2012).

[11] S. Shimizu, K. S. Takahashi, T. Hatano, M. Kawasaki, Y. Tokura, and Y. Iwasa, Phys. Rev. Lett. **111**, 216803 (2013).

[12] R. O. Cherifi, V. Ivanovskaya, L. C. Phillips, A. Zobelli, I. C. Infante, E. Jacquet, V. Garcia, S. Fusil, P. R. Briddon, N. Guiblin, A. Mougin, A. A. Ünal, F. Kronast, S. Valencia, B. Dkhil, A. Barthélémy, and M. Bibes, Nat. Mater. **13**, 345 (2014).

[13] T. Nozaki, A. Kozioł-Rachwał, W. Skowroński, V. Zayets, Y. Shiota, S. Tamaru, H. Kubota, A. Fukushima, S. Yuasa, and Y. Suzuki, Phys. Rev. Appl. **5**, 044006 (2016).

[14] S. Miwa, M. Suzuki, M. Tsujikawa, K. Matsuda, T. Nozaki, K. Tanaka, T. Tsukahara, K. Nawaoka, M. Goto, Y. Kotani, T. Ohkubo, F. Bonell, E. Tamura, K. Hono, T. Nakamura, M. Shirai, S. Yuasa, and Y. Suzuki, Nat. Commun. **8**, 15848 (2017).

[15] C. Takahashi, M. Ogura, and H. Akai, J. Phys. Condens. Matter **19**, 365233 (2007).

[16] M. Oba, K. Nakamura, T. Akiyama, T. Ito, M. Weinert, and A. J. Freeman, Phys. Rev. Lett. **114**, 107202 (2015).

[17] A.-M. Pradipto, T. Akiyama, T. Ito, and K. Nakamura, Phys. Rev. B **96**, 014425 (2017).

[18] F. Ando, H. Kakizakai, T. Koyama, K. Yamada, M. Kawaguchi, S. Kim, K. J. Kim, T. Moriyama, D. Chiba, and T. Ono, Appl. Phys. Lett. **109**, 022401 (2016).

[19] T. Dohi, S. Kanai, A. Okada, F. Matsukura, and H. Ohno, AIP Adv. **6**, 075017 (2016).

[20] T. Dohi, S. Kanai, F. Matsukura, and H. Ohno, Appl. Phys. Lett. **111**, 072403 (2017).

[21] P. F. Carcia, J. Appl. Phys. **63**, 5066 (1988).

[22] C. Kittel, *Introduction to Solid State Physics* (Wiley, New York, 1976).

[23] F. J. Dyson, Phys. Rev. **102**, 1230 (1956).

[24] G. Moulas, A. Lehnert, S. Rusponi, J. Zabloudil, C. Etz, S. Ouazi, M. Etzkorn, P. Bencok, P. Gambardella, P. Weinberger, and H. Brune, Phys. Rev. B **78**, 214424 (2008).

[25] U. Gradmann, Appl. Phys. **3**, 161 (1974).

[26] C. Bi, Y. Liu, T. Newhouse-Illige, M. Xu, M. Rosales, J. W. Freeland, O. Mryasov, S. Zhang, S. G. E. Te Velthuis, and W. G. Wang, Phys. Rev. Lett. **113**, 267202 (2014).

[27] U. Bauer, L. Yao, A. J. Tan, P. Agrawal, S. Emori, H. L. Tuller, S. Van Dijken, and G. S. D. Beach, Nat. Mater. **14**, 174 (2015).

[28] C. Kittel and J. K. Galt, *Ferromagnetic Domain Theory* (1956), pp. 437–564.

[29] L. Onsager, Phys. Rev. **65**, 117 (1944).

[30] S. T. Bramwell and P. C. W. Holdsworth, J. Phys. Condens. Matter **5**, L53 (1993).

[31] M. F. Collins, *Magnetic Critical Scattering* (Oxford Univ. Press, Oxford, New York, 1989).

[32] T. Koyama, A. Obinata, Y. Hibino, A. Hirohata, B. Kuerbanjiang, V. K. Lazarov, and D. Chiba, Appl. Phys. Lett. **106**, (2015).

[33] K. Hüller, J. Magn. Magn. Mater. **61**, 347 (1986).

**Figure Captions**

FIG 1. (a) Experimental arrangement of the devices fabricated for our study. (b) Microscopic image of the device. (c) Schematic of the measurement design. The reduction in magnetization with temperature scales with Bloch's law for thermal spin wave excitation (Eq. (1)). The electric field effect on the exchange coupling $J_{ex}$ can be determined by fitting the temperature dependence of magnetization under various gate voltages $V_G$ with Bloch's law.

FIG 2. (a) Temperature dependence of the normalized perpendicular remanence magnetization $m_\perp$ normalized by $m_\perp$ at 0 K ($=m_S$) with the fittings of Bloch's law. The 3D model and the correction term $\Theta$ (Eq. (1)) agrees well with the result. (b) Temperature dependence of $m_\perp/m_S$ under a gate voltage $V_G = \pm 14$ V. The electric-field-dependent variation of $J_{ex}$ is determined by fitting the results with Eq. (1). (c) Cyclic measurements of the electric field effect on $J_{ex}$.

FIG 3. (a) Temperature dependence of perpendicular component of the magnetization $m_\perp$ per unit area $A$ with a perpendicular magnetic field of 200 Oe. The cross points represent the spontaneous magnetizations $m_{sp}/A$ determined by Arrott plots. The dark green curve

indicates the fitted curve using Eq. (2) to determine $T_C$. (b) Temperature dependence of $m_\perp/A$ under a gate voltage $V_G = \pm 10$ V. The electric-field-dependent variation of $T_C$ is determined by fitting the results with Eq. (2). (c) Cyclic measurements of the electric field effect on $T_C$.

FIG 4. (a) $V_G$ dependence of Curie temperature $T_C$ and exchange coupling $J_{ex}$ normalized by $T_C$ ($V_G = 0$ V) and $J_{ex}$ ($V_G = 0$ V). (b) $J_{ex}$ versus $T_C$ for three different devices. The linear relationship between $J_{ex}$ and $T_C$ follows Weiss molecular field theory.

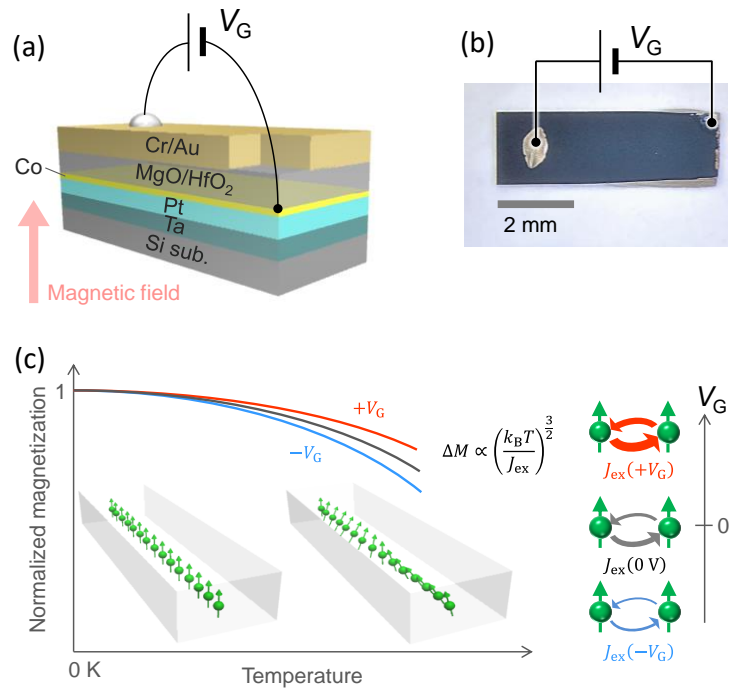

Figure 1

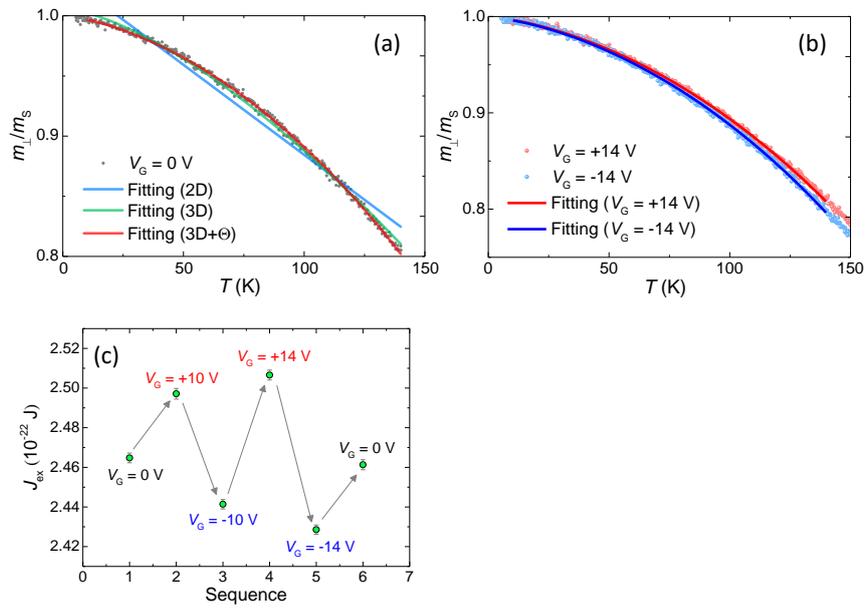

Figure 2

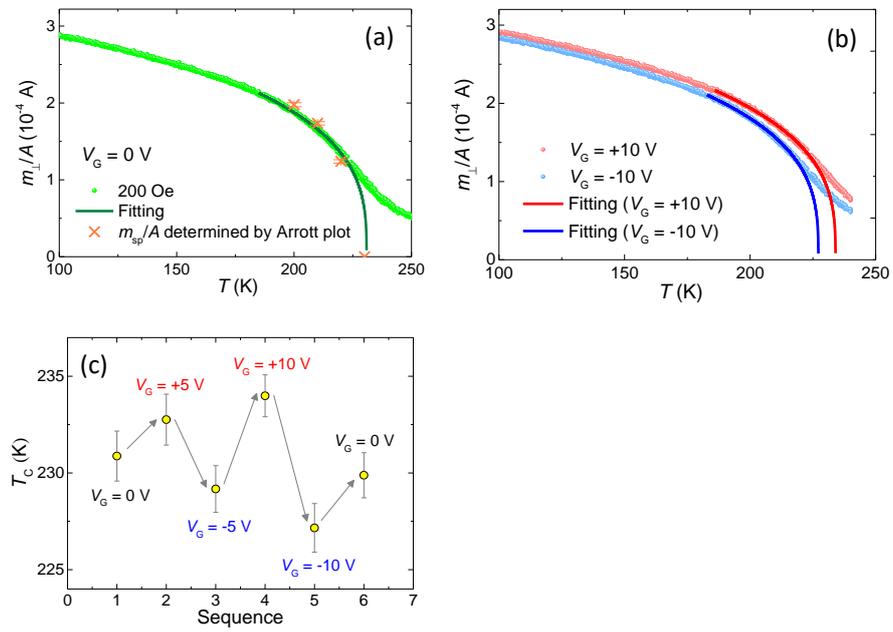

Figure 3

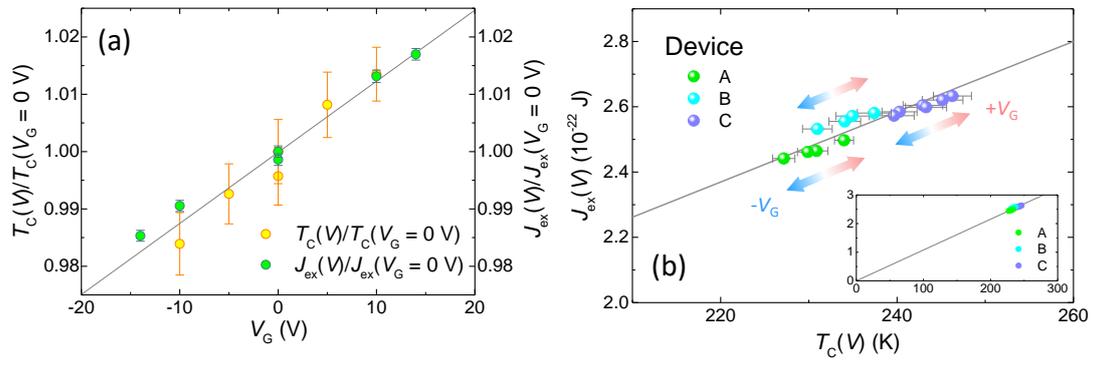

Figure 4

# Microscopic origin of electric-field-induced modulation of Curie temperature in cobalt
-Supplemental Materials-

Fuyuki Ando[1], Kihiro T. Yamada[1], Tomohiro Koyama[2], Mio Ishibashi[1],

Yoichi Shiota[1], Takahiro Moriyama[1], Daichi Chiba[2], and Teruo Ono[1, 3*]

[1] *Institute for Chemical Research, Kyoto University, Gokasho, Uji, Kyoto, 611-0011, Japan*

[2] *Department of Applied Physics, Faculty of Engineering, The University of Tokyo, Hongo 7-3-1, Bunkyo, Tokyo 113-8656, Japan*

[3] *Center for Spintronics Research Network, Graduate School of Engineering Science, Osaka University, Machikaneyama 1-3, Toyonaka, Osaka 560-8531, Japan*

**S1. Device fabrication**

The films of Ta (3.3 nm)/Pt (1.8 nm)/Co (0.23 nm)/MgO (2.0 nm) were prepared by RF sputtering on an intrinsic Si (001) substrate with a thermally oxidized layer ($SiO_2$). The thickness of the layers was determined from the deposition rate of each material. The as-grown film was cut into a size of 2 mm × 5.5 mm in order to set it in a Magnetic Property Measurement Systems (MPMS XL) chamber. Then, the sample was covered by a 40-nm-thick $HfO_2$ gate insulator layer in an atomic layer deposition chamber. Finally, a Cr (3.0 nm)/Au (10 nm) metal electrode was deposited by a resistance heating evaporation system.

**S2. Arrott plots**

The magnetic field dependence of the perpendicular magnetization $m$ per unit area $A$ was measured at various temperatures, as shown in Figure S1(a). The magnetization curve at low temperature (140 K) exhibits an almost full remanence at 0 Oe, indicating that the remanence magnetization $m_\perp$ is equal to the saturation magnetization below that temperature. To determine the spontaneous magnetization near the Curie temperature, an Arrott plot was applied. Figure S1(b) shows the results at various temperatures and the spontaneous magnetizations are determined from the intercept of linear fitting line in the range of 100 Oe $\leq H_\perp \leq$ 400 Oe.

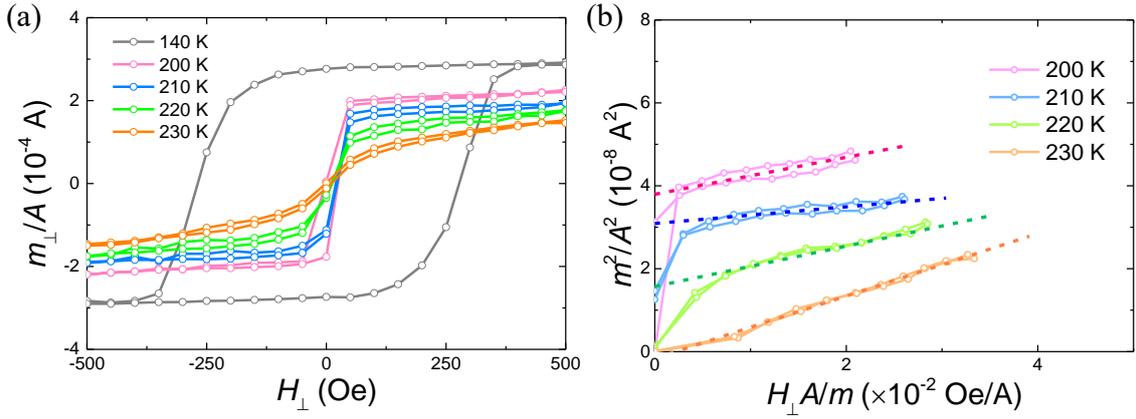

FIG. S1. (a) Magnetic field dependence of the perpendicular magnetization $m$ per unit area $A$ at low temperatures. (b) Arrott plots determined from the magnetization curves at various temperatures. The spontaneous magnetization was determined from the intercept of linear fit.